\documentclass{moriond}

\def\be{\begin{equation}}
\def\ee{\end{equation}}
\def\bea{\begin{eqnarray}}
\def\eea{\end{eqnarray}}

\newcommand{\Photo}{}

\begin{document}
\vspace*{4cm}
\title{Quarkonium and charm at Belle II}

\author{ Michel Bertemes on behalf of the Belle II Collaboration }

\address{Brookhaven National Laboratory, Physics Department\\
11973 Upton (NY), USA}

\maketitle\abstracts{
We present recent results from the Belle II experiment related to quarkonium and charm physics. With data samples collected by Belle II during operation of the SuperKEKB collider above the $\Upsilon(4S)$ resonance, we study the processes $e^+e^- \to \omega\chi_{bJ}(1P)$ ($J=0$, 1, or 2), search for the bottomonium equivalent of the X(3872) and measure the exclusive cross sections of $e^+e^- \to B\bar{B}$, $B\bar{B}^*$ and $B^*\bar{B}^*$. We also propose a new algorithm to determine the production flavor of neutral $D$ mesons. Finally, we report on a measurement of the $\Omega_c^0$ lifetime. }

\section{Quarkonium}

\subsection{Observation of $e^+e^-\to \omega \chi_{bJ}(1P)$ and search for $X_b\to \omega \Upsilon(1S)$ at $\sqrt{s}$ near 10.75 GeV}

Heavy quarkonium is well suited to study the non-perturbative behavior of quantum chromodynamics. The $\Upsilon(10753)$ is one of four bottomonium-like vector states identified above the $B\bar{B}$ threshold \cite{Workman:2022ynf}. It was observed by the Belle experiment in $e^+e^-\to \pi^+\pi^- \Upsilon(nS)$ ($n=1,2,3$) \cite{Belle:2019cbt} and in fits to the $e^+e^-\to b\bar{b}$ cross sections at energies $\sqrt{s}$ from 10.6 to 11.2 GeV \cite{Dong:2020tdw}. One of many interpretations sees the $\Upsilon(10753)$ as an admixture of $\Upsilon(4S)$ and  $\Upsilon(3D)$ \cite{Li:2021jjt} with branching fractions for decays into $\omega \chi_{bJ}$ (where $\chi_{bJ}$ denotes $\chi_{bJ}(1P)$) of $10^{-3}$, comparable to the decay rate via di-pion transitions. This final state also allows for the search of the process $e^+e^-\to \gamma X_b(\to \omega \Upsilon(1S))$, where $X_b$ is the supposed bottomonium counterpart of the $X(3872)$. Previous searches for an $X_b$ state by ATLAS \cite{ATLAS:2014mka} , CMS \cite{CMS:2013ygz} and Belle \cite{Belle:2014sys} observed no signal. For these analyses we use data samples collected by the Belle II experiment at four different centre-of-mass energies above the $\Upsilon(4S)$: $\sqrt{s}=$ 10.653, 10.701, 10.745 and 10.805 GeV with integrated luminosities of 3.5, 1.6, 9.8 and 4.7 $\mathrm{fb}^{-1}$, respectively. 

\begin{figure}
    \centering
    \includegraphics[width=0.7\textwidth]{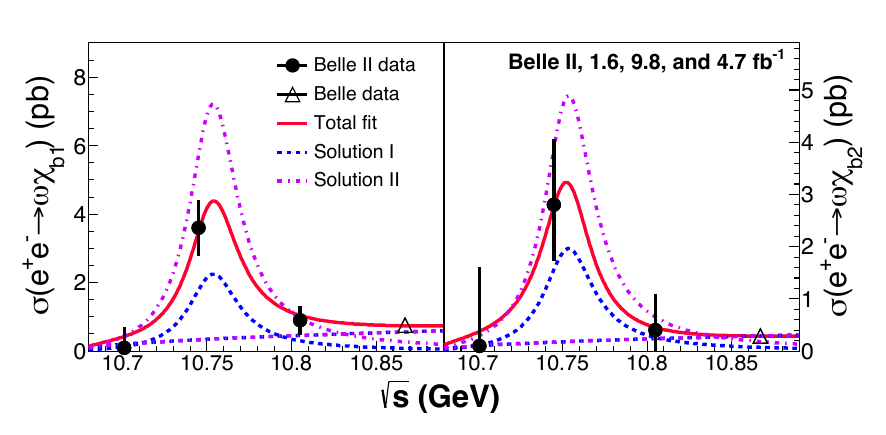}
    \caption{Energy dependence of the Born cross sections with fit results overlaid for (left) $e^+e^- \to \omega \chi_{b1}$ and (right) $e^+e^- \to \omega \chi_{b2}$. Belle II results from this study are indicated with circles, while previous measurements by Belle \protect\cite{Belle:2014sys} are shown with triangles. Solution I (II) represents constructive (destructive) interference between $\omega \chi_{b1}$ and $\omega \chi_{b2}$.}
    \label{fig:upsilon}
\end{figure}

We select events containing at least 4 tracks to search for with $\omega \to \pi^+\pi^-\pi^0$, $\chi_{bJ}\to\gamma \Upsilon(1S)$, and $\Upsilon(1S)\to \ell^+\ell^-$ ($\ell$=$e$ or $\mu$). We perform two-dimensional unbinned likelihood fits to the $M(\gamma\Upsilon(1S))$ versus $M(\pi^+\pi^-\pi^0)$ distributions and find signals of $\chi_{b1}$ and $\chi_{b2}$ at 10.745 and 10.805 GeV with significances of $11\sigma$ and $4.5\sigma$, respectively. This marks the first observation of $\omega\chi_{bJ}$ signals at $\sqrt{s}=$10.745 GeV. We measure the corresponding Born cross sections to be $(3.6^{+0.7}_{-0.7}(\mathrm{stat})\pm 0.5 (\mathrm{syst}))\,$pb and $(2.8^{+1.2}_{-1.0}(\mathrm{stat})\pm 0.4 (\mathrm{syst}))\,$pb. A strong enhancement of the cross section is observed near 10.75 GeV with an energy dependence consistent with the $\Upsilon(10753)$ state (see Figure \ref{fig:upsilon}).

In distributions of $M(\omega\Upsilon(1S))$, we find reflections from $e^+e^-\to \omega \chi_{bJ}$ signals but no narrow peak corresponding to a $X_b$ signal. We therefore determine upper limits at 90\% Bayesian credibility on the products of Born cross section for $e^+e^-\to \gamma X_b$ and branching fraction for $X_b\to\omega \Upsilon(1S)$ to be 0.55, 0.84, 0.14 and 0.37 pb at $\sqrt{s}=$10.653, 10.701, 10.745 and 10.805 GeV, respectively.  

\subsection{Measurement of energy dependence of the $e^+e^- \to B\bar{B}, B\bar{B}^*$ and $B^*\bar{B}^*$ cross sections} 

The energy region surrounding the previously discussed $\Upsilon(10753)$ suffers from great uncertainty as the available data samples are limited in size. When Belle II collected its first scan data, the center-of-mass energies were chosen to fill the gaps in between the scan points of Belle and therefore improve the understanding of the concerned region. We report the measurement of the $e^+e^-\to B\bar{B}$, $B\bar{B}^*$ and $B^*\bar{B}^*$ cross sections using the data samples collected at 10.653, 10.701, 10.745 and 10.805 GeV, following a previously established procedure \cite{Belle:2021lzm}. One of the two $B$ mesons in the event is reconstructed in decays to hadronic final states using the Full Event Interpretation package of Belle II \cite{Keck:2018lcd}. We identify the different $B\bar{B}$, $B\bar{B}^*$ and $B^*\bar{B}^*$ signals with the $M_{\mathrm{bc}}$ distribution:
\begin{equation}
        M_{\mathrm{bc}} = \sqrt{ (E_{\mathrm{cm}}/2)^2 - p_B^2 }
\end{equation}
which represents the invariant mass of the reconstructed $B$ meson with the energy replaced by half the centre-of-mass energy $E_{\mathrm{cm}}$, and $p_B$ being the $B$ candidate momentum in the centre-of-mass frame. Signal yields are obtained from fits to the $M_{\mathrm{bc}}$ distributions at various scan energies. We compute the corresponding Born cross sections and find them in good agreement with earlier measurements from Belle \cite{Belle:2021lzm} with better precision. Finally, we perform a simultaneous fit of the energy dependence of the exclusive $e^+e^-\to B\bar{B}$, $B\bar{B}^*$ and $B^*\bar{B}^*$ cross sections from Belle and Belle II together with the total $b\bar{b}$ cross section \cite{Dong:2020tdw}. The sum of exclusive cross sections agrees well with the total one, which represents an important cross check as different methods are used to obtain both. We also see a sharp increase at the $B^*\bar{B}^*$ threshold which could hint at a possible resonance. Further studies are needed to investigate this observation. 
   
\section{Charm}

\subsection{Novel method for the identification of the production flavor of neutral charmed mesons}

One of the main ingredients of any $C\!P$ violation and mixing measurement is the identification of the signal flavor at production, referred to as flavor tagging. In charm physics, this is usually accomplished by selecting $D^0$ mesons from the strong decay $D^{*+}\to D^0 \pi^+$ or from the semi-leptonic decay of a beauty hadron. The sample of neutral $D$ mesons available for measurements that require tagging is therefore much smaller than the inclusive sample of neutral $D$ mesons produced in $e^+e^-$ collisions.

We present a new approach to charm-flavor tagging (Charm Flavor Tagger, CFT) that exploits the correlation between the flavor of the signal $D$ meson and the electric charges, flavor and kinematic properties of particles reconstructed in the rest of the $e^+e^-\to c\bar{c}$ event. The latter includes particles from the decay of the other charm hadron in the event and those possibly produced in association with the signal meson. The CFT selects particles most collinear with the signal meson as these are most likely to correctly tag the flavor. It uses a binary classification algorithm to predict the product $qr$, where $q$ is the tagging decision ($q=+1$ for $D^0$ and $q=-1$ for $\bar{D}^0$) and $r$ is the dilution ($r=0$ indicates that the flavor is not known, while $r=1$ corresponds to a perfect prediction). The input variables are a set of reconstructed quantities of the selected particles related to kinematics and particle identification discriminators.

\begin{figure}
    \centering
    \includegraphics[width=0.5\textwidth]{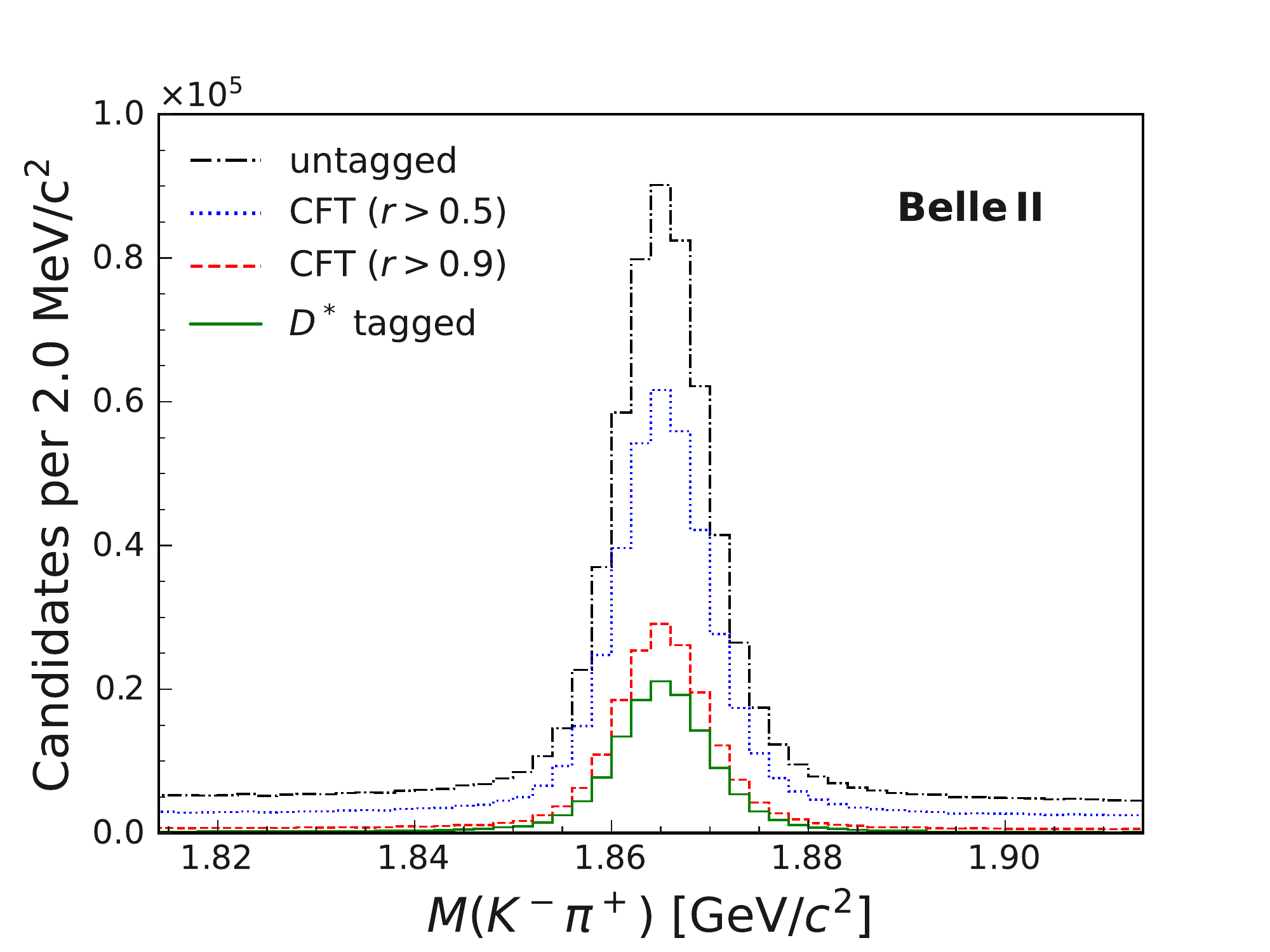}
    \caption{Mass distribution for $D^0\to K^- \pi^+$ decays with different requirements on the predicted dilution in comparison with $D^{*+}$-tagged decays.}
    \label{fig:cft_kpi}
\end{figure}

The CFT is trained using simulation and calibrated with data collected by Belle II corresponding to an integrated luminosity of 362$\mathrm{fb}^{-1}$. The tagging power, defined as $\epsilon_{\mathrm{tag}}^{\mathrm{eff}}=\epsilon_{\mathrm{eff}}\langle r^2\rangle$ ($\epsilon_{\mathrm{eff}}$ being the tagging efficiency), represents the effective sample size when a tagging decision is needed. We obtain:
\begin{equation}
    \epsilon_{\mathrm{tag}}^{\mathrm{eff}} = (47.91\pm0.07(\mathrm{stat})\pm0.51(\mathrm{syst}))\%
\end{equation}
This value is found to be independent of the signal neutral-$D$ decay mode. The CFT will roughly double the effective sample size with respect to measurements that so far have relied exclusively on $D^{*+}$ tagged events. The increase in sample size is accompanied by an increase in background (see Figure \ref{fig:cft_kpi}).

\subsection{Measurement of the $\Omega_c^0$ lifetime}

The lifetime hierarchy of heavy hadrons is predicted in the context of heavy-quark expansion while expressing the decay rate as an expansion in inverse powers of the heavy-quark mass. For charmed hadrons, an accurate prediction is challenging as higher-order terms in $1/m_c$ and contributions of spectator quarks cannot be neglected, leading to overall large uncertainties \cite{Lenz:2014jha}. According to early measurements and in agreement with theoretical predictions, the $\Omega_c^0$ was believed to be the shortest lived among the four singly charmed baryons that decay weakly. We measure the $\Omega_c^0$ lifetime using $\Omega_c^0 \to \Omega^- \pi^+$ decays with $\Omega ^- \to \Lambda^0(\to p\pi^-) K^-$ and data collected by Belle II corresponding to 207 $\mathrm{fb}^{-1}$ to be: 
\begin{equation}
    \tau(\Omega_c^0) = (243\pm48(\mathrm{stat})\pm11(\mathrm{syst})) \mathrm{fs}
\end{equation}
This result provides an independent experimental confirmation of earlier determinations by the LHCb experiment \cite{LHCb:2018nfa,LHCb:2021vll} which challenged the existing world average value and set the $\Omega_c^0$ as the second-longest living charmed baryon. With this program of charm lifetime measurements, we continue to demonstrate the excellent performance and alignment of the vertex detector.

\section*{Acknowledgments}

The author would like to thank the Belle II Collaboration for the opportunity to present, and the organizers of the 57th Rencontres de Moriond for the successful conference. This work is supported by the Austrian Science Fund under award number J4625-N.

\end{document}